\documentclass{PoS}
\usepackage{graphicx,wrapfig}
\usepackage{epsfig}
\usepackage{placeins}

\def\l14{$L_{\rm 1.4GHz}$}
\def\s14{$S_{\rm 1.4GHz}$}

\def\emerlin{$e$-MERLIN}
\def\emerge{$e$MERGE}

\newcommand{\etal}{et~al.}

\def\gs{\mathrel{\raise0.35ex\hbox{$\scriptstyle >$}\kern-0.6em\lower0.40ex\hbox{{$\scriptstyle \sim$}}}} 
\def\ls{\mathrel{\raise0.35ex\hbox{$\scriptstyle <$}\kern-0.6em\lower0.40ex\hbox{{$\scriptstyle \sim$}}}}

\def\Wm2{\,\hbox{W}\,\hbox{m}^{-2}} 
\def\gsim{\mathrel{\raise0.35ex\hbox{$\scriptstyle >$}\kern-0.6em\lower0.40ex\hbox{{$\scriptstyle \sim$}}}} 
\def\lsim{\mathrel{\raise0.35ex\hbox{$\scriptstyle <$}\kern-0.6em\lower0.40ex\hbox{{$\scriptstyle \sim$}}}} 
\def\ltsima{$\; \buildrel < \over \sim \;$} 
\def\simlt{\lower.5ex\hbox{\ltsima}} 
\def\gtsima{$\; \buildrel > \over \sim \;$} 
\def\simgt{\lower.5ex\hbox{\gtsima}}

\title{Studying galaxy evolution through cosmic time via the $\mu$Jy radio population: early results from $e$MERGE}

\ShortTitle{$e$MERGE Data Release 1}

\author{\speaker{A.\,P.\ Thomson}$^{a,b}$, T.\,W.\,B.\ Muxlow$^{a}$, Ian Smail$^{c}$, I.\,M.\ McHardy$^{b}$, R.\,J.\ Beswick$^{a}$, J.\,F.\,Radcliffe$^{a,d,e}$, N.\ Wrigley$^{a}$, on behalf of the \emerge\ consortium\\
\llap{$^a$}Jodrell Bank Centre for Astrophysics, The University of Manchester, Oxford road, Manchester, M13 9PL, UK\\
\llap{$^b$}Department of Physics and Astronomy, The University of Southampton, Southampton, SO17 1BJ, UK\\
\llap{$^c$}Centre for Extragalactic Astronomy, Department of Physics, Durham University, South Road, Durham, DH1 3LE, UK\\
\llap{$^d$}Kapteyn Astronomical Institute, University of Groningen, 9747 AD Groningen, NL\\
\llap{$^e$}ASTRON, the Netherlands Institute for Radio Astronomy, Postbus 2, 7990 AA Dwingeloo, NL\\
\email{alasdair.thomson@manchester.ac.uk}}

\abstract{The \emerlin\ Galaxy Evolution Survey (\emerge) is an ambitious, multi-tiered extragalactic radio continuum survey being carried out with \emerlin\ and the VLA at 1.4\,GHz and 6\,GHz. Exploiting the unique combination of high sensitivity and high angular resolution provided by radio interferometry, these observations will provide a powerful, obscuration-independent tool for tracing intense star-formation and AGN activity in galaxies out to $z\sim5$. In our first data release (DR1) we present \emerge\ Tier 1, a 15-arcmin pointing centred on the GOODS-N field, imaged at 1.4\,GHz with the VLA and \emerlin\ at $\sim 0.28''$ resolution down to an rms sensitivity of $\sim 1.2\,\mu$Jy\,beam$^{-1}$. This unique radio survey -- unrivaled at 1.4\,GHz in its combination of depth, areal coverage and angular resolution in the pre-SKA era -- allows us to localise and separate extended star-forming regions, nuclear starbursts and compact AGN core/jet systems in galaxies over the past two-thirds of cosmic history, a crucial step in tracing the apparently simultaneous growths of the stellar populations and central black holes in massive galaxies.

In these proceedings we highlight some early science results from \emerge\ DR1, including some examples of the sub-arcsecond morphologies and cold dust properties of 1.4\,GHz-selected galaxies. \emerge\ Tier 1 will eventually reach sub-$\mu$Jy\,beam$^{-1}$ sensitivity at $0.28''$ resolution over a 30-arcmin field, providing crucial benchmarks for deep extragalactic surveys which will be undertaken with SKA in the next decade.}
\FullConference{14th European VLBI Network Symposium \& Users Meeting (EVN 2018)\\
		8-11 October 2018\\
		Granada, Spain}

\begin{document}

\section{Introduction}
\vspace*{-4mm}
One of the key goals of modern cosmology is to understand the formation of the whole galaxy population through cosmic time. Historically, optical and near-infrared surveys have served as the key workhorses for measuring the integrated star-formation history of the Universe, however over the past two decades it has been demonstrated that a pan-chromatic (i.e.\ X-ray -- radio) approach is key to achieving a consensus view on galaxy evolution (e.g.\ Driver \etal, 2009). Since early work in the far-IR/sub-mm with SCUBA, it has been established that a significant fraction of the star-formation through cosmic history (up to $\sim 50\%$ at $z=1$--$3$; Swinbank \etal, 2014) occurred in heavily dust-obscured galaxies, which are typically faint (or undetected) in even the deepest optical/IR studies (e.g. Casey \etal, 2014, and references therein). Deep interferometric radio continuum observations provide an indispensible, dust-unbiased tracer of star formation (e.g. Smolcic \etal, 2017), complementing studies in the optical/IR, and allowing us to track the build-up of stellar populations through cosmic time. Moreover, radio continuum observations also provide a direct probe of active galactic nuclei (AGN), which are believed to play a crucial role in the evolution of their host galaxies via feedback effects (e.g. Harrison \etal, 2015).

\begin{wrapfigure}{r}{7.5cm}
\includegraphics[width=7.5cm]{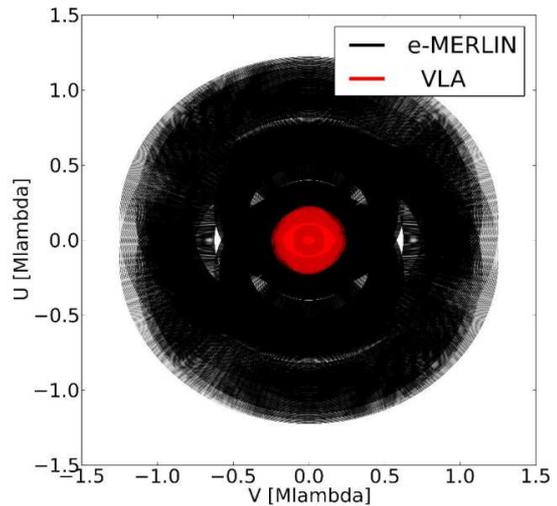}
\caption{$uv$ coverage of our combined-array \emerge\ 1.4\,GHz dataset. \emerlin's long baselines (up to $\sim 200$\,km) hugely extend the VLA-only $uv$ coverage, while the presence of short spacings from the VLA (0.68--27\,km) fill the ``hole'' in the $uv$ plane which is present in \emerlin-only observations (due to a shortest baseline length of $\sim 10$\,km). The resulting $uv$ coverage of this combined dataset provides sensitive imaging on scales from $0.28''$--$40''$.}\label{fig:uvcoverage}
\end{wrapfigure} 

However, the very fact that radio continuum emission is produced by both star-formation and AGN activity makes the interpretation of radio observations of distant galaxies non-trivial. To differentiate between these two processes we require either (a) the angular resolution and surface brightness sensitivity to decompose (extended) star-formation from (point-like) nuclear activity and jet structures (e.g. Baldi \etal, 2018; Jarvis \etal, \textit{submitted}), or (b) multi-frequency observations which provide the spectral index information necessary to reliably isolate ``radio excess'' sources, i.e. galaxies whose radio luminosities exceed those which can be attributed to star-formation (e.g. Magnelli \etal, 2014; Thomson \etal, 2014). To spatially-resolve massive galaxies at $z\gtrsim 1$ and create maps of star-formation and AGN activity requires sensitive ($\sigma_{\rm rms}\sim 1\,\mu$Jy\,beam$^{-1}$) imaging with $\sim$kpc (i.e. sub-arcsecond) resolution. While the VLA is currently capable of achieving this combination of observing goals with deep integrations between $S$ and $X$-band (Smolcic \etal, 2017; Murphy \etal, 2017, respectively), at $z\gtrsim 2$ these observations probe rest-frame frequencies of $\nu_{\rm rest}\sim 10$--$30$\,GHz -- a portion of the radio spectrum in which spectral curvature becomes important due to the increasing contribution of thermal free-free emission (e.g. Murphy \etal, 2011). This complicates efforts to measure the rest-frame radio luminosities ($L_{\rm 1.4\,GHz}$) of galaxies at high-redshift. Moreover, at $10$\,GHz, the instantaneous field of view (FoV) of the VLA is small ($\theta_{\rm PB}\sim 4.5'$, $cf.$ $\theta_{\rm PB}\sim 32'$ at 1.4\,GHz), while at 1.4\,GHz the FoV is large, but the VLA lacks the resolving power to morphologically study the bulk of the high-redshift galaxy population ($\theta_{\rm res}\sim 1.5''$, corresponding to $\sim 10$\,kpc at $z\sim 2$).

\begin{figure*}
\centerline{\psfig{file=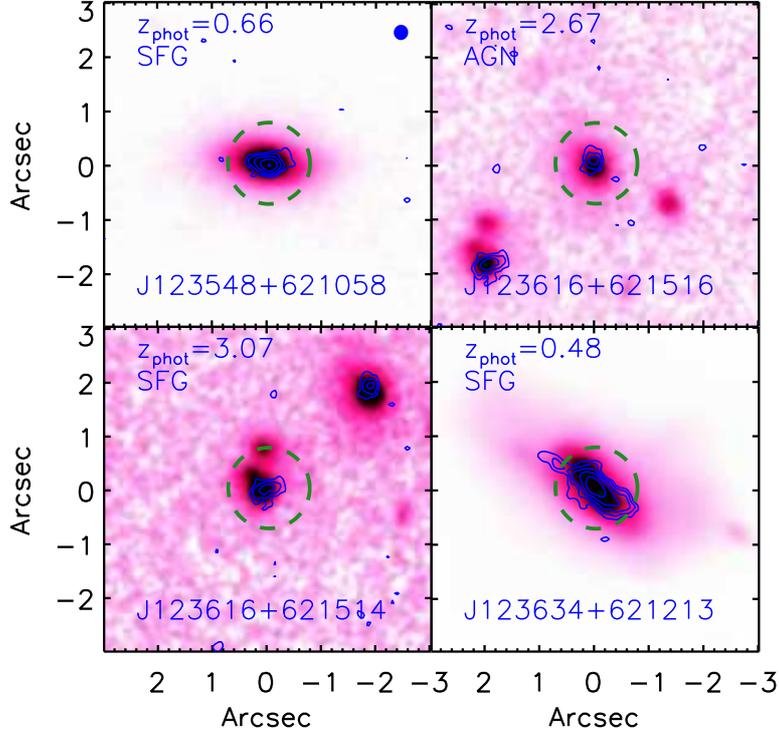,width=10cm}}
\caption[Short captions]{Example thumbnail images of four galaxies from our 1.4\,GHz \emerlin +VLA combined-array imaging of GOODS-N. The naturally-weighted radio maps are shown as blue contours (spaced at $-3$, $3$, $3\times\sqrt{2}\sigma$ and in steps of $\sqrt{2}\times\sigma$ thereafter, with $\sigma\sim 1.2\,\mu$Jy\,beam$^{-1}$), while the underlying heat-map is the \textit{HST} WFC3 F125W image from the CANDELS legacy programme (Grogin \etal, 2011; Koekemoer \etal, 2011). The green dashed circle has a {\sc fwhm} of $1.5''$, equivalent to the VLA 1.4\,GHz A-array PSF, and highlights the improvement in angular resolution offered by our combined-array image over the VLA alone. For comparison, the $0.28''$ PSF of our naturally-weighted \emerlin +VLA image is shown in blue in the top-right of the first sub-figure. Galaxies are classified as either star-forming galaxies or AGN-dominated systems based on their ratios of far-infrared ($8$--$1000\,\mu$m) and radio luminosities (see \S\,\ref{sect:qir}).}
\label{fig:stamps}
\end{figure*}

To address these issues, we have been conducting the \emerlin\ Galaxy Evolution Survey (\emerge; Muxlow \etal. 2019, \textit{in prep}), an ambitious legacy survey which combines observations of the GOODS-N extragalactic field at 1.4\,GHz and 6\,GHz from both \emerlin\ and the Karl G.\, Jansky Very Large Array (VLA). The long baselines ($B_{\rm max}\sim 200$\,km) and large collecting area of \emerlin ($\sim 1/3$ of the total collecting area of SKA-1) provide an excellent complement to the high surface brightness sensitivity of the VLA, allowing the production of radio maps which exceed the capabilities of either telescope individually.

\section{\emerlin\ \& VLA data reduction and imaging}
\vspace*{-4mm}
We here summarize the data processing and imaging strategies for our \emerlin\ and VLA data. Further details will be provided in the forthcoming \emerge\ survey description paper (Muxlow \etal, 2019, \textit{in prep}), while details of the construction of the radio source catalogue will be provided in Thomson \etal, 2019 (\textit{in prep}).

\FloatBarrier
\begin{wrapfigure}{L}{7.5cm}
\includegraphics[width=7.5cm]{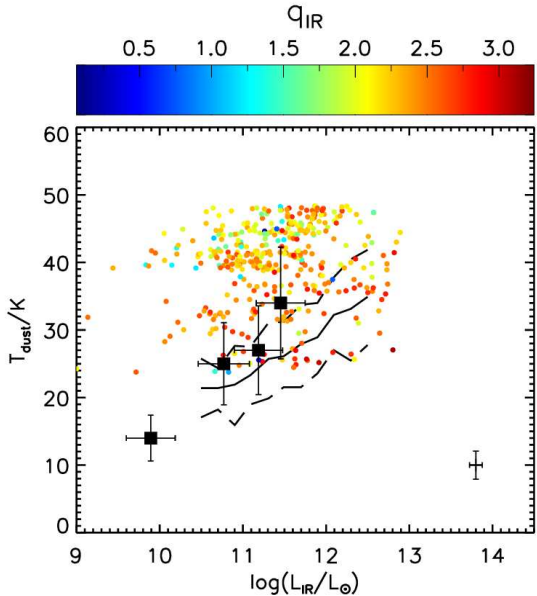}
\caption{Dust temperature ($T_{\rm dust}$) versus $8$--$1000\,\mu$m luminosity ($L_{\rm IR}$) for the 378 radio-selected \emerge\ galaxies with deblended \textit{Herschel} SPIRE photometry. Points are colour-coded by the logarithmic ratio of far-IR/radio flux, $q_{\rm IR}$. We also show the relation for $0.1<z<2$ IR-selected star-forming galaxies from Symeonidis \etal\ (2013; black solid line with $\pm2\sigma$ scatter in the relation shown with dashed lines), and stacked H$\alpha$-selected star-forming galaxies from the HiZELS survey (Thomson \etal, 2017; filled squares). A representative error bar for the \emerge\ sample is shown in the bottom-right corner. We find that at a given $L_{\rm IR}$, our radio-selected sample has warmer dust emission compared to galaxies selected in the far-IR, in good agreement with previous studies (e.g. Casey \etal, 2009, Amblard \etal, 2010). Not shown are the $\sim 240$ galaxies for which only upper-limits to their deblended far-IR photometry are available, which likely includes a significant population of radio-excess AGN.}\label{fig:lirtd}
\end{wrapfigure} 
\FloatBarrier

Our \emerge\ DR1 image comprises $\sim 180$\,hours of \emerlin\ L-band data (the first $\sim 1/4$ of the total time awarded to the project) plus $\sim 30$\ hours from the VLA in A-array (previously published by Owen, 2018). Together, these two datasets offer baselines from $\sim 0.68$\,km--$200$\,km, providing near-complete $uv$ coverage and sensitivity to emission on scales from the angular resolution of \emerlin\ ($\theta_{\rm res}\sim 0.2''$) up to the largest angular scale of the VLA ($\theta_{\rm LAS}\sim 40''$; Fig.\,\ref{fig:uvcoverage}).

We processed our \emerlin\ data manually using {\sc aips} version {\sc 31dec14}, including one \textit{a priori} round of automated flagging of bad data using the {\sc serpent} tool (Peck \& Fenech, 2013), followed by extensive manual flagging using the {\sc aips} tasks {\sc ibled} and {\sc spflg}. The data were calibrated using standard techniques (e.g. Wrigley, 2016), with one round of phase-only self-calibration performed on the target field prior to imaging. We processed our VLA data by performing one round of automated flagging using the {\sc aoflagger} package (Offringa \etal, 2012) then used the scripted {\sc casa} pipeline (version 4.3.1) to perform further automated flagging as well as flux density, phase and bandpass calibration. On completion of the {\sc casa} pipeline tasks, one round of phase-only self calibration was performed using a model of the inner $5$-arcmin of the field, before the data were reweighted for optimal sensitivity using the {\sc casa} task {\sc statwt}.

The final calibrated datasets from both arrays were written to {\sc casa} Measurement Set format, along with re-processed legacy datasets from the pre-upgrade VLA and MERLIN arrays (Muxlow \etal, 2005). A naturally-weighted map was produced from this combined dataset using the {\sc wsclean} (version 2.6; Offringa \etal, 2014) package, which utilises the $w$-stacking approach to mitigate the effects of the non-coplanar array geometry. Our DR1 image covers the inner 15-arcmin of the GOODS-N field with a pixel scale of $0.045''$ (i.e. a $20$\,k $\times 20$\,k pixel image), is cleaned using an auto-masking procedure, and reaches an rms sensitivity of $\sigma_{\rm rms}\sim 1.2\,\mu$Jy\,beam$^{-1}$ near the pointing centre. We show a selection of thumbnails from this radio map in Fig.\,\ref{fig:stamps}.

\vspace*{-3mm}
\section{Star-forming galaxies and AGN at high-redshift}\label{sect:qir}
\vspace*{-4mm}

The radio continuum spectra of normal star-forming galaxies and AGN are typically comprised of power-law components (tracing synchrotron and thermal free-free emission) and lack the spectral features required to measure accurate redshifts. To allow meaningful astrophysics to be derived from our radio source catalogue, we therefore cross-match our \emerge\ source list with the multiwavelength (optical--mid-IR) photometric catalogue of GOODS-N compiled by the 3D-HST team (Skelton \etal, 2014), which includes photometric redshifts calculated using the {\sc eazy} code (Brammer \etal, 2008). We find 3D-HST optical counterparts to 612/820 \emerge\ sources within our DR1 image, with the majority of the non-matched radio sources lying outside the footprint of the \textit{Hubble} WFC3 mosaic image used to define the 3D-HST survey area.

We supplement this multiwavelength coverage with far-IR flux density mesaurements from the \textit{Herschel} PACS ($100$ \& $160\,\mu$m) and SPIRE ($250$--$500\,\mu$m) imaging of GOODS-N (Elbaz \etal, 2011). The angular resolution of \textit{Herschel} SPIRE is $\sim 15$--$30''$, significantly under-sampling the density of galaxies in the \emerge\ catalogue. To measure reliable far-infrared flux densities for individual galaxies we therefore need to deblend the SPIRE maps for the effects of this confusion. Following the techniques of Swinbank \etal\ (2014) and Thomson \etal\ (2017), we exploit the extensive multi-wavelength imaging of GOODS-N to construct a catalogue of infrared and radio-bright galaxies which can be used as positional priors to deblend the SPIRE maps (removing any sources which lie within $<1.5''$ as duplicates), and insert these priors as delta functions with arbitrary flux densities into blank model maps matching the astrometry and pixel scale of the original SPIRE images. Next, we use a Monte Carlo algorithm to converge on the optimum SPIRE flux densities for these sources by randomising the flux densities of the delta functions, convolving this model with the appropriate SPIRE PSF and searching for the set of fluxes which minimises the residual image (i.e. data--model).

We measure far-infrared ($8$--$1000\,\mu$m) luminosities, $L_{\rm IR}$, and dust temperatures, $T_{\rm dust}$, for 378 galaxies (of the 612 with photometric redshifts) in the \emerge\ sample by fitting model spectral energy distributions from a library which includes high-redshift dust-obscured starburst and AGN templates (see Thomson \etal, 2017 for details) to the deblended far-IR photometry. The dust temperatures of these radio-selected galaxies span a range $T_{\rm dust}\sim 25$-$50$\,{\sc k} (Fig.\,\ref{fig:lirtd}), a little higher than those of IR-selected galaxies at the same $L_{\rm IR}$ (e.g. Symeonidis \etal, 2013). From these far-IR luminosities we also measure $q_{\rm IR}\equiv \log_{10}[(S_{\rm IR}/3.75\times 10^{12}{\rm W\,m}^{-2})/(S_{\rm 1.4\,GHz}/{\rm W\,m}^{-2}{\rm Hz}^{-1})]$, following Ivison \etal\ (2010), where $S_{\rm 1.4\,GHz}$ is the rest-frame 1.4\,GHz flux density ($k$-corrected assuming $S_{\nu}\propto \nu^\alpha$ with $\alpha=-0.8$) and $S_{\rm IR}$ is the rest-frame $8$--$1000\,\mu$m flux. This logarithmic flux ratio provides a clear discriminator between galaxies whose radio luminosities are consistent with star-formation ($q_{\rm IR}\sim 2.2$) and those whose radio luminosities can only be explained via the presence of a powerful AGN ($q_{\rm IR}\lesssim 1.68$; Del Moro \etal, 2013).

\section{Conclusions}
\vspace*{-4mm}
We have presented example thumbnails and some early science results from our \emerlin + VLA 1.4\,GHz imaging survey of the GOODS-N field, constituting Data Release 1 of Tier 1 of the \emerlin\ Galaxy Evolution Survey. By combining $\sim 10$--$200$\,km baselines from \emerlin\ with $<30$\,km baselines from the VLA in A-configuration and performing a single imaging run on the gridded $uv$ data, we achieve a ``best of both worlds'' imaging scenario, offering sensitivity on scales from the $\sim 0.28''$ PSF up to the largest angular scale of the VLA, $\theta_{\rm LAS}\sim 40''$. In combination with published optical--mid-IR data and new deblended \textit{Herschel} SPIRE photometry, this ambitious legacy programme promises to provide new insights into the population of radio-selected star-forming galaxies and AGN from $z\sim 1$--$5$.

\FloatBarrier
\section{Acknowledgements}
\vspace*{-4mm}
APT acknowledges support from STFC (ST/P000649/1). \emerlin\ is a National Facility operated by the University of Manchester at Jodrell Bank Observatory on behalf of STFC. The National Radio Astronomy Observatory is a facility of the National Science Foundation operated under cooperative agreement by Associated Universities, Inc.

\end{document}